\documentclass[aps,twocolumn,prl,reprint,amsmath,amssymb,showpacs,superscriptaddress]{revtex4-1}

\usepackage{graphicx}
\usepackage{dcolumn}
\usepackage{bm}
\usepackage{amsmath}
\usepackage{color}
\usepackage{mathrsfs}
\usepackage{float}
\usepackage{dsfont}
\usepackage{txfonts}
\usepackage{wasysym}
\usepackage{ifsym}

\makeatletter

\newcommand{\Rmnum}[1]{\expandafter\@slowromancap\romannumeral #1@}
\makeatother
\begin{document}

\title{Enhanced thermoelectric response in the fractional quantum Hall effect}   
    
\author{Pablo Roura-Bas}
\affiliation{Dpto de F\'{\i}sica, Centro At\'{o}mico Constituyentes, Comisi\'{o}n
Nacional de Energ\'{\i}a At\'{o}mica, CONICET, Buenos Aires, Argentina}

\author{Liliana Arrachea} 
\affiliation{International Center for Advanced Studies, ECyT  Universidad Nacional de San Mart\'{\i}n, 
Campus Miguelete, 25 de Mayo y Francia, 1650 Buenos Aires, Argentina}
\affiliation{Dahlem Center for Complex Quantum Systems and Fachbereich Physik, Freie Universit\"at Berlin, 14195 Berlin, Germany}

\author{Eduardo Fradkin}
\affiliation{Department of Physics and Institute for Condensed Matter Theory,
University of Illinois at Urbana-Champaign, 1110 West Green Street, Urbana, Illinois 61801-3080, USA}

\begin{abstract}
We study the linear thermoelectric response of a quantum dot embedded in a constriction of a quantum Hall bar with fractional filling factors $\nu=1/m$ within Laughlin series. We calculate the figure of merit $ZT$ 
for the maximum efficiency at a fixed temperature difference. 
We find a significant enhancement of this quantity in the fractional filling in relation to the integer-filling case, which is a direct consequence of the fractionalization of the electron in the fractional quantum Hall state. We present simple theoretical expressions for the Onsager coefficients at low temperatures, which explicitly show that
$ZT$ and the Seebeck coefficient increase with $m$. 
\end{abstract}

\date{\today}

\maketitle


{\em Introduction}
Boosting the efficiency for the conversion of electrical and thermal energy at finite power is motivating an intense research activity, not only in the areas of material science and applied physics but also in experimental and
theoretical areas of 
statistical mechanics and condensed matter physics.   Efforts are concentrated  on developing new materials and devices \cite{thermomat} as well as on analyzing different operational conditions
\cite{thermocond}.
In the latter direction, taking advantage of the quantum effects is one of the most interesting avenues. 
Nanostructures operating at low temperatures are particularly appealing quantum devices, since they offer the conditions for coherent
transport, where ``parasitic'' heat currents by phonons are strongly suppressed. Quantum dots (QD) are one of the most studied nanostructures in this context. Due to their spectrum of discrete levels, amenable to be manipulated 
with gate voltages, they can be used as switches for the relevant transport channels. Theoretically, they were found to present high thermoelectric response  \cite{thermoqd1,thermoqd2,thermoqd3,thermoqd4,thermoqd5,thermoqd6,thermoqd7}. 

A two dimensional electron gas in the  quantum Hall effect (QHE) regime hosts chiral edges states \cite{wen,chang,eduardo, laug, Halperin-1982, but}. Due to their topological protection, these states constitute the paradigmatic system 
to realize the
coherent transport regime with fractionalized excitations.  
Electron transport through QDs in QHE structures was studied in Ref. \onlinecite{furusaki,kim}. The usefulness of the thermoelectric transport in these structures to enable the detection of neutral modes in fractional fillings $\nu=2/3$ 
and $\nu=5/2$ 
was analyzed in Refs. \onlinecite{stern,heiblum}.  The nature of the thermal transport  in the QHE has been investigated theoretically \cite{ k-f-1, k-f-2, cappelli, grosfeld, us, us1} and experimentally 
\cite{granger, Nam, altimiras, yacoby, altimiras2, qcond,baner} in integer and fractional fillings. Thermoelectric effects induced by  interferences by multiple quantum point contacts in fractional fillings were studied in 
Ref. \onlinecite{vanuci}. However, the thermoelectric performance has been
so far investigated only 
within the integer QHE beyond linear response \cite{rosa} and 
in multiterminal systems \cite{rafa}. In the last case, the possibility of a separating  heat and charge currents provides a route to improving the thermoelectric performance. The fact that the partitioning of charge and energy
are not trivially related in tunneling junctions between Luttinger liquids was pointed out in Ref. \cite{torsten}.
The goal of the present work is to show 
that the fractionalization of the charge  also offers a mechanism for thermoelectric enhancement, which manifests itself even in a simple two-terminal configurations of QHE systems.

\begin{figure}[tbp]
\centerline{
\includegraphics[width=\columnwidth]{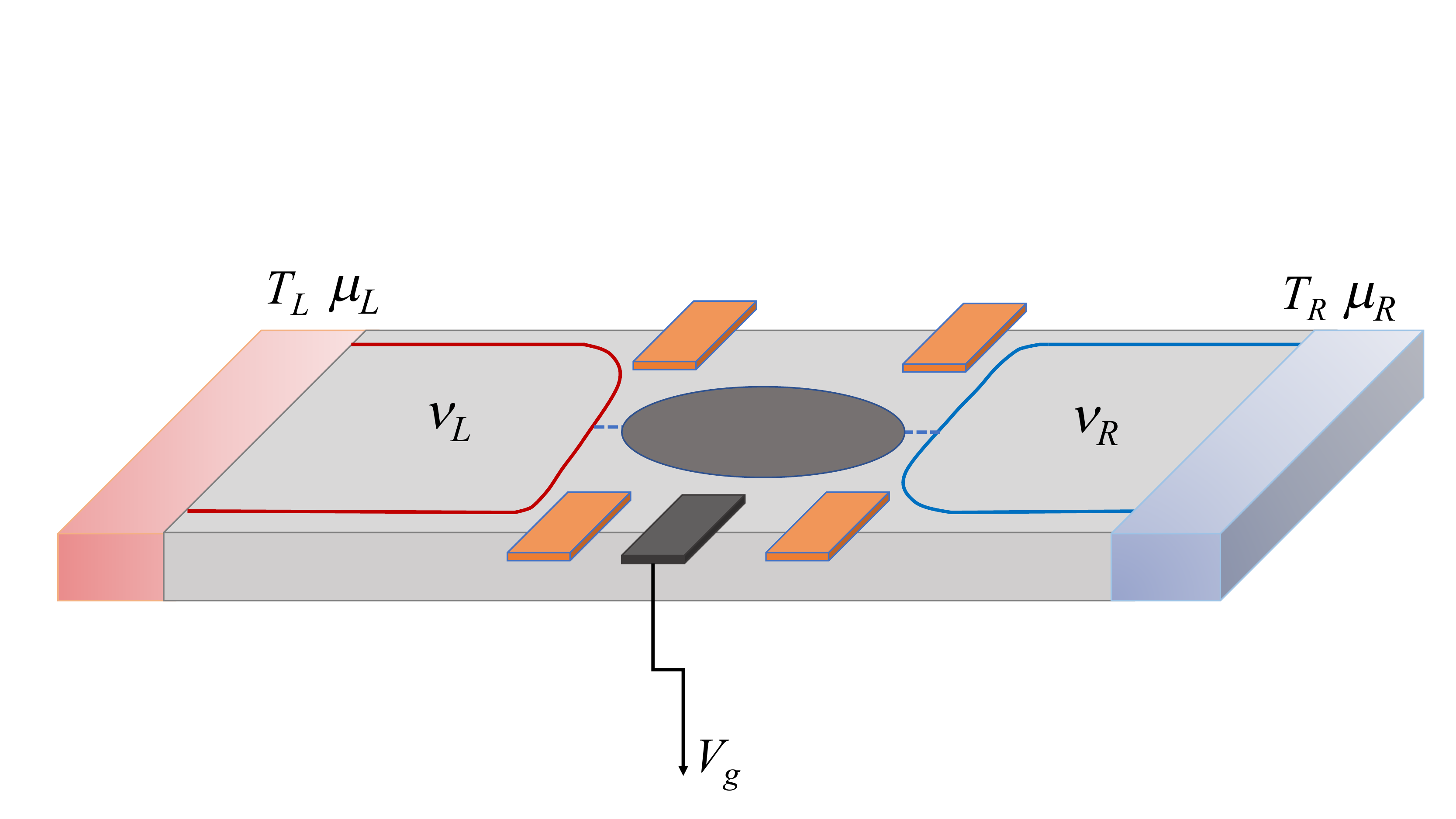}}
\caption{(Color online) Sketch of the setup. A quantum Hall bar is biased by a difference of temperature $\Delta T=T_L-T_R$ and a voltage $V=(\mu_L-\mu_R)/e$. Charge and thermal transport is induced through a 
quantum dot generated by constrictions that generate regions of the sample with filling factors $\nu_L$ and $\nu_R$. The spectrum of the quantum dot can be manipulated with a gate voltage $V_g$.} 
\label{fig1}
\end{figure}

We analyze the thermoelectric efficiency of QHE structures, focusing on fractional fillings within the Laughlin series $\nu=1/m$.  We consider the setup
 sketched in Fig. \ref{fig1}, where a QD is embedded into a constriction of a  QHE bar containing regions with filling factors $\nu_L$ and $\nu_R$. 
 The QD
 is contacted to the corresponding edge states through quantum point contacts.
 Electric and heat currents, respectively denoted by  $J_C$ and $J_Q$ flow through the quantum dot as a response to  chemical potential 
 and temperature biases applied at the contacts, $\mu_R-\mu_L= eV$
  and $\Delta T = T_L- T_R$, respectively.   The device may operate
 as a heat engine, in which case the efficiency 
is defined as the ratio $\eta_{\rm he}=P/J_Q$, between the generated power $P=J_C V$ and the heat current from the hot to the cold reservoir. The other operational mode is a refrigerator, which is characterized by a coefficient of performance
$\eta_{\rm fri}= J_Q/P$, where $J_Q$ is
the heat current extracted from the cold reservoir and $P$ is the invested electrical power. Both coefficients are bounded by the Carnot values.
In the linear response regime, relevant for small $V$ and $\Delta T$, $\eta_{\rm he, fri}$ can be parametrized by the ``figure of merit'' $ZT$ \cite{iofe}, defined below, in a way that $ZT \rightarrow \infty$ implies the Carnot limit. 
Remarkably, we will show that $ZT$ is significantly enhanced in the fractional quantum Hall effect, relative to the response in the integer one. This enhancement is a direct consequence of the 
fractionalization of the electron in the fractional QHE.

{\em Linear thermoelectric response.}
Following the conventions of Ref. \onlinecite{casati}, we consider $ \Delta T= T_L - T_R >0 $ and $ \mu_L -\mu_R = eV  < 0$. The two relevant fluxes are the charge and heat currents, which are represented by the vector 
$\bm{ J} \equiv \left( J_C, \; J_Q \right)$ and corresponds to the charge and heat current leaving the left  contact of the bar. In linear response they are related to  the affinities represented by the vector  
$\bm{ X} = \left(eV/k_B T, \Delta  T/k_B T^2 \right)$ through the Onsager matrix $\hat{L}$ as  $\bm{  J} = \hat{L} \; \bm{ X}$. The diagonal matrix elements of $\hat{L}$ are related to the electrical and thermal conductivity, while the off diagonal ones are related to the Seebeck and Peltier coefficients.  These four coefficients characterize the transport properties of the device. In the presence of an external magnetic field, the off-diagonal ones
 obey Onsager reciprocity relations $L_{12}(B) = L_{21}(-B)$. Due to the symmetry of the two-terminal setup we are considering they also satisfy $L_{12}(B)=L_{21}(B)$. In addition, the second law of thermodynamics implies 
 $L_{11}, L_{22} \geq 0$ and $\mbox{det}[\hat{L}] >0$. Taking into account these conditions, it can be shown 
 that the maximum achievable efficiency in any of the two operational modes at a fixed temperature difference can be parametrized in terms of the figure of merit $ZT=L_{12}^2/\mbox{det}[\hat{L}]$
as follows \cite{casati}
\begin{equation}\label{eta}
\eta_{\rm he, fri}^{max}= \eta_{\rm he,fri}^C \frac{\sqrt{ZT+1}-1}{\sqrt{ZT+1}+1}.
\end{equation}
 $\eta^C_{\rm he}= 1 - T_R/T_L$ and $\eta^C_{\rm fri}= T_L/(T_L - T_R)$ are 
the Carnot efficiency and coefficient of performance
for the heat-engine and the  refrigerator, respectively. They set a bound for the coefficient of Eq. (\ref{eta}), which is achieved when $ZT \rightarrow \infty$. 

{\em Model and currents.} We consider the following Hamiltonian for the full setup $H=\sum_{\alpha=L,R}H_{\alpha}+H_d+H_t$.
The first term represents the edge states of the QHE with filling factors $\nu_{\alpha}= 1/m_{\alpha}$, which is described by the  following Hamiltonian
\begin{equation}\label{hal}
H_{\alpha}=\frac{\pi {\rm v}}{\nu_{\alpha}}\int dx~\rho^{2}_{\alpha}(x),
\end{equation}
where $\alpha=L,R$ denotes propagating modes injected from the left and right contacts and moving along the edge with velocity ${\rm v}$. The corresponding densities are 
$\rho_{\alpha}(x)=\partial_x \phi_{\alpha}(x)/(2 \pi)$ where $\phi_{\alpha}(x)$ are chiral bosonic fields  that satisfy the Kac-Moody algebra
$\left[\phi_{\alpha}(x),\phi_{\beta}(x^{\prime}) \right]  = -i \pi \nu_{\alpha} \delta_{\alpha,\beta} \mbox{sg}(x-x^{\prime})$. 
The second term of the Hamiltonian $H$ describes the QD. It reads $H_d = \sum_{j=1}^N \varepsilon_{d,j} d^{\dagger}_j d_j$, where we are assuming a large Zeeman term that justifies considering  fully polarized electrons.  
We assume that the $N$ levels of the quantum dot are equally spaced in energy by $\Delta$ and can be shifted  by recourse to the gate voltage 
as $\varepsilon_{d_j}=\Delta (j-1) - e V_g$. The third term of the Hamiltonian represents the tunneling between the edge states and the QD,
\begin{equation} \label{ht}
H_{t} = {\cal V}_t \sum_{j, \alpha =L, R} \left[\Psi^{\dagger}_{ \alpha }(x_0) d_{j}+ H. c. \right],
\end{equation}
with ${\cal V}_t= \sqrt{2 \pi a} V_t$, being $V_t$ the tunneling kinetic energy and $a$ is a characteristic length setting the high energy cutoff for the edge spectrum, while  $x_0$ denotes the position of the edge at which the  contact to the dot is established.  The bosonic form of the electron operator at the edge is  \cite{eduardo}
\begin{equation}\label{bosonization2}
\Psi_{\alpha}(x)\equiv  \frac{F_{\alpha}}{\sqrt{2\pi a}}~e^{i\frac{s_{\alpha}}{\nu_{\alpha}} \phi_{\alpha}(x,t)},
\end{equation}
where $F_{\alpha}$ are Klein factors.

According to our definitions the charge and heat currents are $J_C= - e \langle \dot{N}_L \rangle$ and $J_Q= J_E - \mu_L J_C $, with  
$J_E=- \langle \dot{H}_L \rangle$. For the model under consideration, we have 
\begin{eqnarray}
J_C &=& \langle \hat{J}_{C} \rangle =  ieV_{t} \langle ~\Psi^{\dagger}_{L}(x_0)d  - d^{\dagger} \Psi_{L}(x_0) \rangle, \nonumber \\
J_E &=& \frac{\pi v}{e\nu_L} \langle ~\hat{J}_{C}\rho_{L}(x_0)+\rho_{L}(x_0) \hat{J}_{C}~\rangle.
\end{eqnarray}
We resort to Schwinger-Keldysh non-equilibrium Green functions to calculate these currents,  starting from their representation in terms the lesser Green functions
$G^{<}_{\alpha, d}(t-t^{\prime})=i\langle d^{\dagger}_{\sigma}(t^{\prime})\Psi_{\alpha}(xt) \rangle$ as follows
\begin{eqnarray}\label{currents}
J_C &=& eV_{t}\Bigr( G^{<}_{d, L }(t-t')-G^{<}_{L, d}(t-t') \Bigl)\vert_{t=t'} , \nonumber \\
J_E&= & iV_{t}\partial_{t}\Bigr( G^{<}_{d, L}(t-t')-G^{<}_{L, d}(t-t') \Bigl)\vert_{t=t'}.              
\end{eqnarray}                           
These expressions can be calculated by recourse to perturbation theory in the tunneling term.  Details are provided in Refs. \onlinecite{sup,martin,bra}. The result is
\begin{eqnarray}\label{jpert}
J_C &=& \frac{e}{h} \int d \varepsilon \; \tau(\varepsilon)\;  \left[ f_L(\varepsilon+\mu_L)- f_R(\varepsilon+\mu_R) \right], \nonumber \\
J_E &=&  \frac{1}{h}\int d \varepsilon \;  \varepsilon \; \tau(\varepsilon) \; \left[ f_L(\varepsilon+\mu_L)- f_R(\varepsilon+\mu_R) \right],
\end{eqnarray}
where $f_{\alpha}(\varepsilon) = 1/\left(e^{\varepsilon/(k_B T_{\alpha})} + 1\right)$ is the Fermi function. We have introduced the transmission function $\tau(\varepsilon)$ defined as
\begin{equation} \label{tau}
\tau(\varepsilon)= \frac{V_t^4}{\gamma} \; D_R(\varepsilon + \mu_R) \; D_d(\varepsilon)\;  D_L(\varepsilon + \mu_L).
\end{equation}
Here,   $D_{\nu}(\varepsilon)$, with $\nu=R, L, d$,  are the density of states of the right, left QHE edge states and the quantum dot, respectively. For the latter, we consider a model
of $N$ resonances with energies $\varepsilon_{d,j}$ and widths $\gamma$.  We assume that  $\gamma \sim V_t$.  The corresponding expressions are
\begin{eqnarray}
D_{\alpha}(\varepsilon)& = &  \frac{a^{m_{\alpha} -1}}{2 \pi} \frac{ (2 \pi T_{\alpha})^{m_{\alpha} -1} }{ \Gamma(m_{\alpha})} 
\left| \frac{\Gamma \left( m_{\alpha}/2 + i \varepsilon /(2 \pi T_{\alpha}) \right) }{\Gamma \left( 1/2 + i \varepsilon /(2 \pi T_{\alpha}) \right)} \right|^2 \nonumber \\
D_d(\varepsilon) & = & \sum_j \frac{\gamma/N\pi}{\left(\varepsilon- \varepsilon_{d,j} \right)^2 + \gamma^2},
\end{eqnarray}
where $\Gamma(z)$ is the Euler function. 

\begin{figure}[tbp]
\centerline{
\includegraphics[width=1.1\columnwidth]{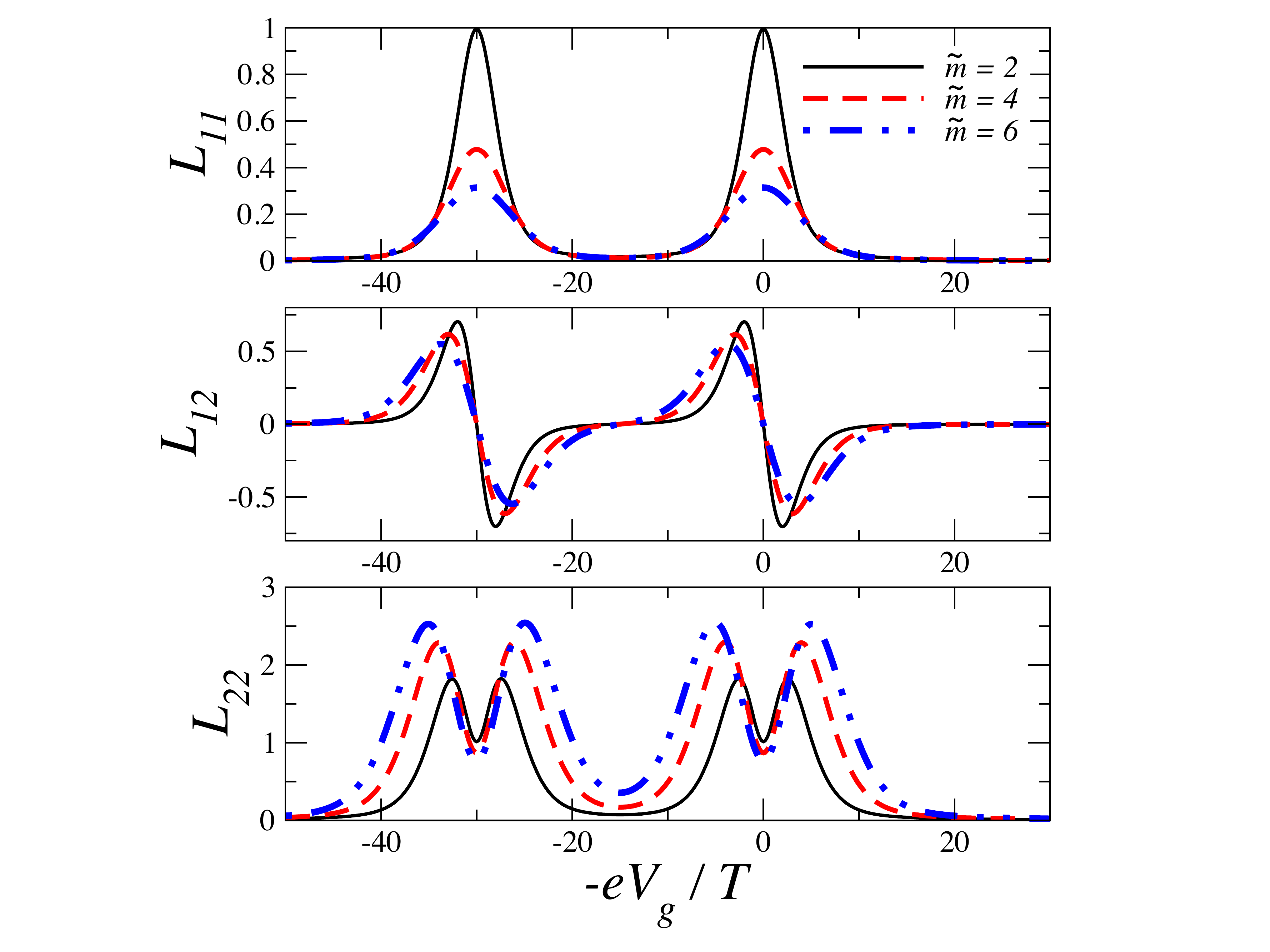}}
\caption{(Color online) Onsager coefficients at  the temperature $T=\gamma$, as functions of the gate voltage for a quantum dot with two energy of the levels. Solid, dashed and dashed-dotted lines correspond to $\tilde{m}=m_L+m_R=2, 4, 6$, respectively. The integer-filling case corresponds to $\tilde{m}=2$. The level spacing is 
$\Delta=30 \gamma$.  The chemical potential is set at $\mu=0$. } 
\label{fig2}
\end{figure}

{\em Onsager matrix and transport coefficients}. 
We take as references $T=T_R$ and 
$\mu=(\mu_L+\mu_R)/2 =0$.  
Expanding the difference of Fermi functions in Eqs. (\ref{jpert}), we have
\begin{equation} \label{l}
\hat{L}=-\frac{k_B T}{2 h} \int d \varepsilon \left(\begin{array}{cc} 
e & e \varepsilon \\
\varepsilon & \varepsilon^2 \end{array} \right) \tau(\varepsilon) \; \frac{\partial f(\varepsilon)}{\partial \varepsilon}.
\end{equation}
The behavior of the different matrix elements is determined by the transmission function $\tau(\varepsilon)$. The latter
can be externally modified by changing the energy of the levels in the quantum dot by means of the gate voltage. The temperature enters the density of states of the edges and the
derivative of the Fermi function. 
Examples  are shown in Fig. \ref{fig2} where the three different $L_{ij}$ are shown as functions of the gate voltage $V_g$ for a given value of the background temperature. We recall that the $V_g$ shifts 
rigidly the energy of the QD levels.
Interestingly, the behavior of these  coefficients is determined by the integer $\tilde{m}=m_L+m_R$
corresponding to an effective filling factor $1/\tilde{\nu}= 1/\nu_L + 1/\nu_R$. This
property was previously discussed in the context of the conductance \cite{claudio1} and the charge-current noise \cite{claudio2} in quantum point contacts between QHE regions with different filling factors.
Here, we see that it is a more general characteristic of all the transport coefficients, which becomes explicit in the analytic low-temperature behavior given
by Eqs. (\ref{lowt}).
The element $L_{11}$ exhibits peaks when a level is aligned with the chemical potential and as a function
of the gate voltage. Instead, the off-diagonal coefficient $L_{12}$ vanishes and changes sign at this value, which indicates the possibility of operating the device as a heat engine ($L_{12} >0$) or a refrigerator ($L_{12}<0$). 
The vanishing of $L_{12}$ implies a lack of thermoelectric response when the chemical potential is exactly resonant with the levels of the dot.  
As a function of the filling fraction, we see that  $L_{11}$ and
$L_{12}$ decrease for the fractional fillings $\tilde{m}>2$ in comparison to the 
integer case $\tilde{m}=2$. 
The suppression of  the electrical conductance in the tunneling regime of the fractional QHE, and in Luttinger liquids in general,  has been widely discussed in the literature \cite{chang}. Here, we see that a similar effect takes place 
in the off-diagonal Onsager coefficient as well.

\begin{figure}[tbp]
\centerline{
\includegraphics[width=0.8\columnwidth]{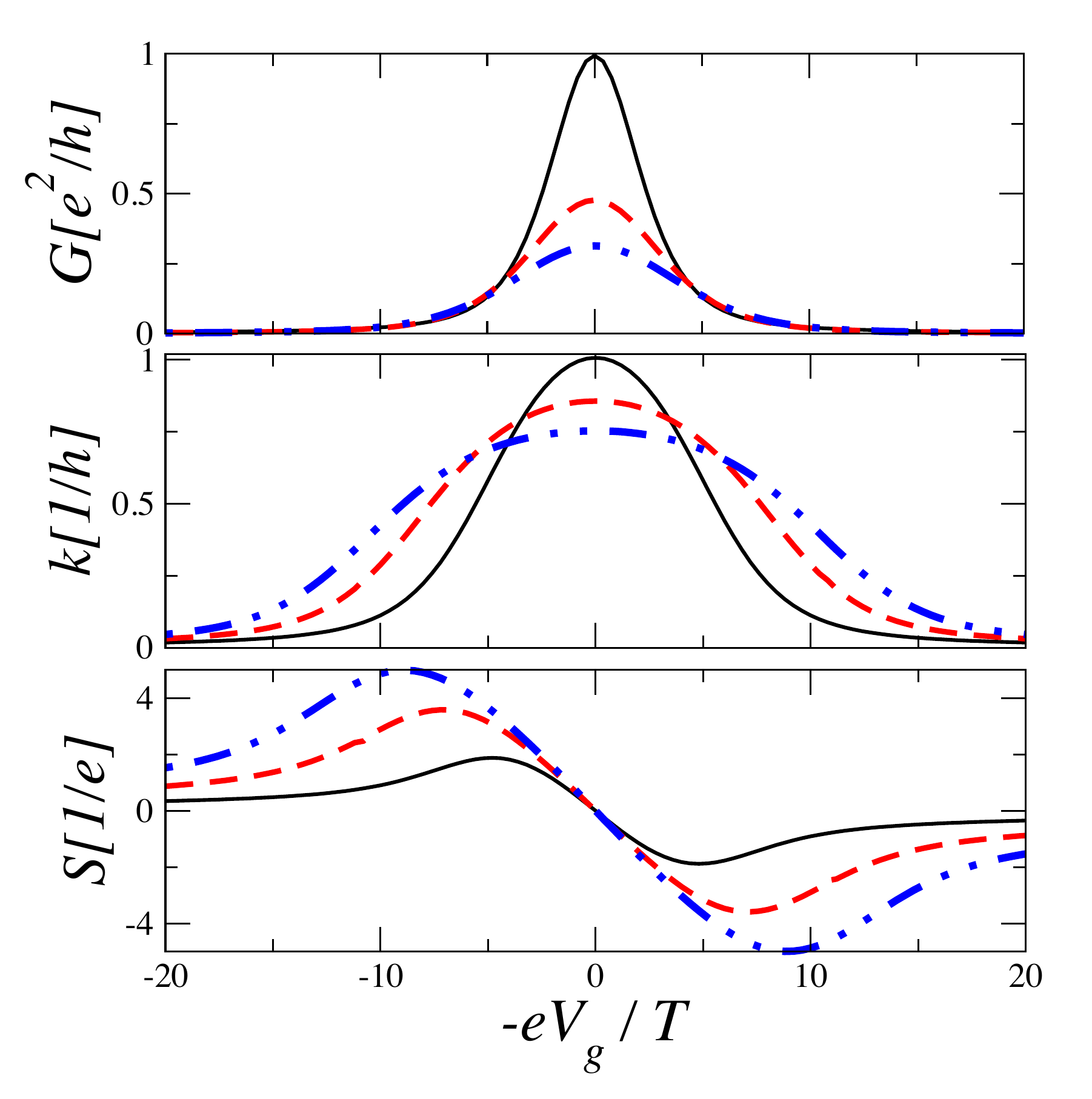}}
\caption{(Color online) Electrical and thermal conductances, $G$ and $\kappa$, respectively and Seebeck coefficient $S$ of a single-level quantum dot, as functions of the gate voltage.
 Other details are the same as in Fig. \ref{fig2}.} 
\label{fig3}
\end{figure}
The Onsager matrix elements are related to  the electrical and thermal conductances $G$ and $\kappa$, the Seebeck and Peltier coefficients $S$ and $\Pi$, 
respectively, as follows
\begin{eqnarray}
G &= & \frac{L_{11}}{T}, \;\;\;\;\;\; \kappa= \frac{1}{T^2} \frac{\mbox{det} \hat{L}}{L_{11}}, \;\;\;\;\;\; T S  = \Pi= \frac{L_{12}}{L_{11}}.
\end{eqnarray}
Examples of their dependence with $V_g$ are shown in Fig. \ref{fig3} for a single-level. We see that the electrical and thermal conductances follow a behavior similar to that of the diagonal Onsager coefficients. Remarkably, we see that the Seebeck coefficient increases in the fractional-filling case, relative to the integer-filling one. This behavior is ``a priori'' unexpected from the behavior of $L_{12}$, which follows the opposite trend. The behavior of $S$ is due to the fact 
that $L_{12}$  decreases as function of the the filling factor at a lower rate than $L_{11}$, as a consequence of  the different impact that the fractionalization of the charge ($m_L, \; m_R$) has on the charge and thermal channels. This feature becomes explicit in the low-temperature behavior described by Eqs. (\ref{lowt}).

\begin{figure}[tbp]
\centerline{
\includegraphics[width=0.5 \columnwidth]{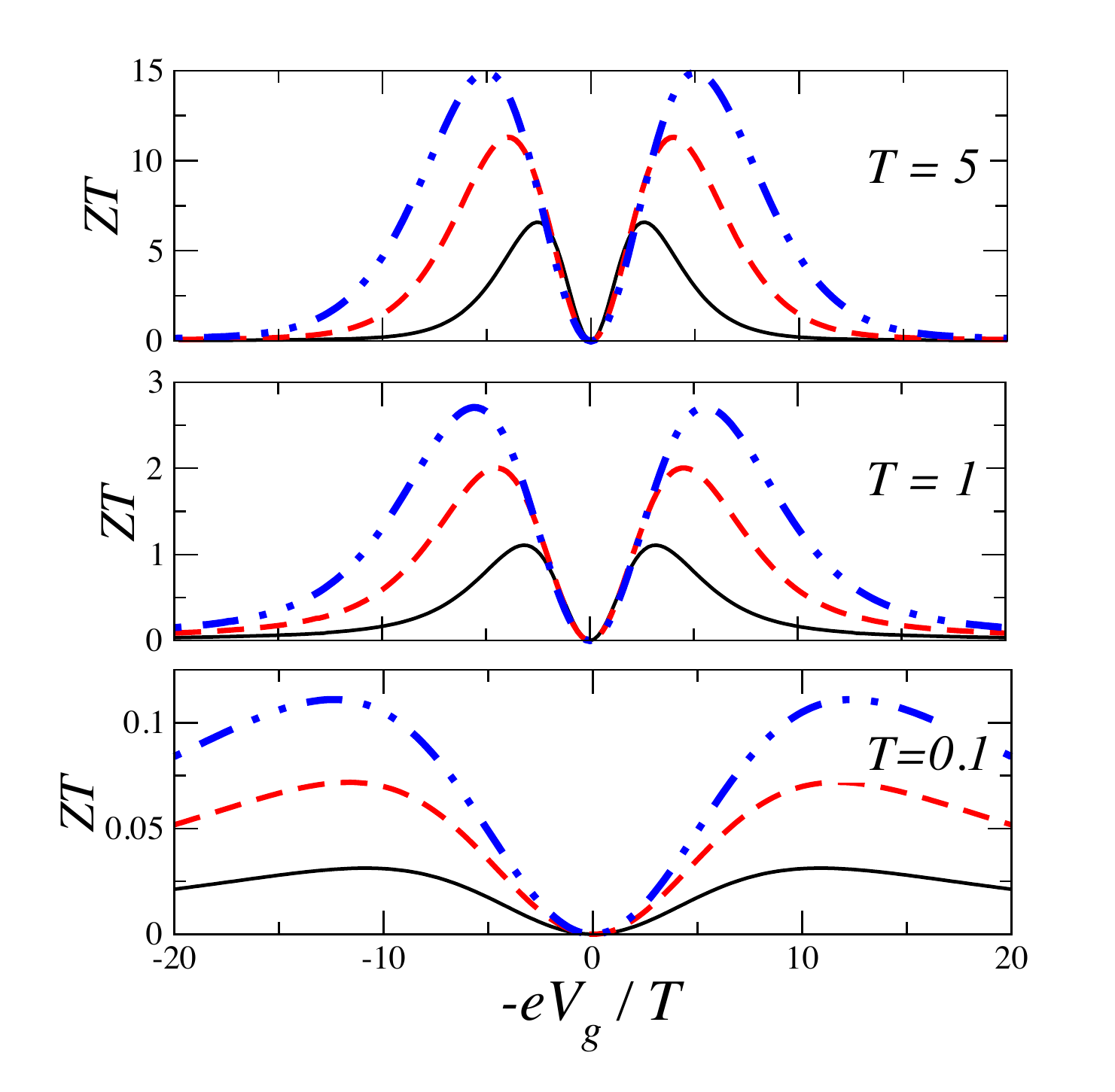}
\includegraphics[width=0.6\columnwidth]{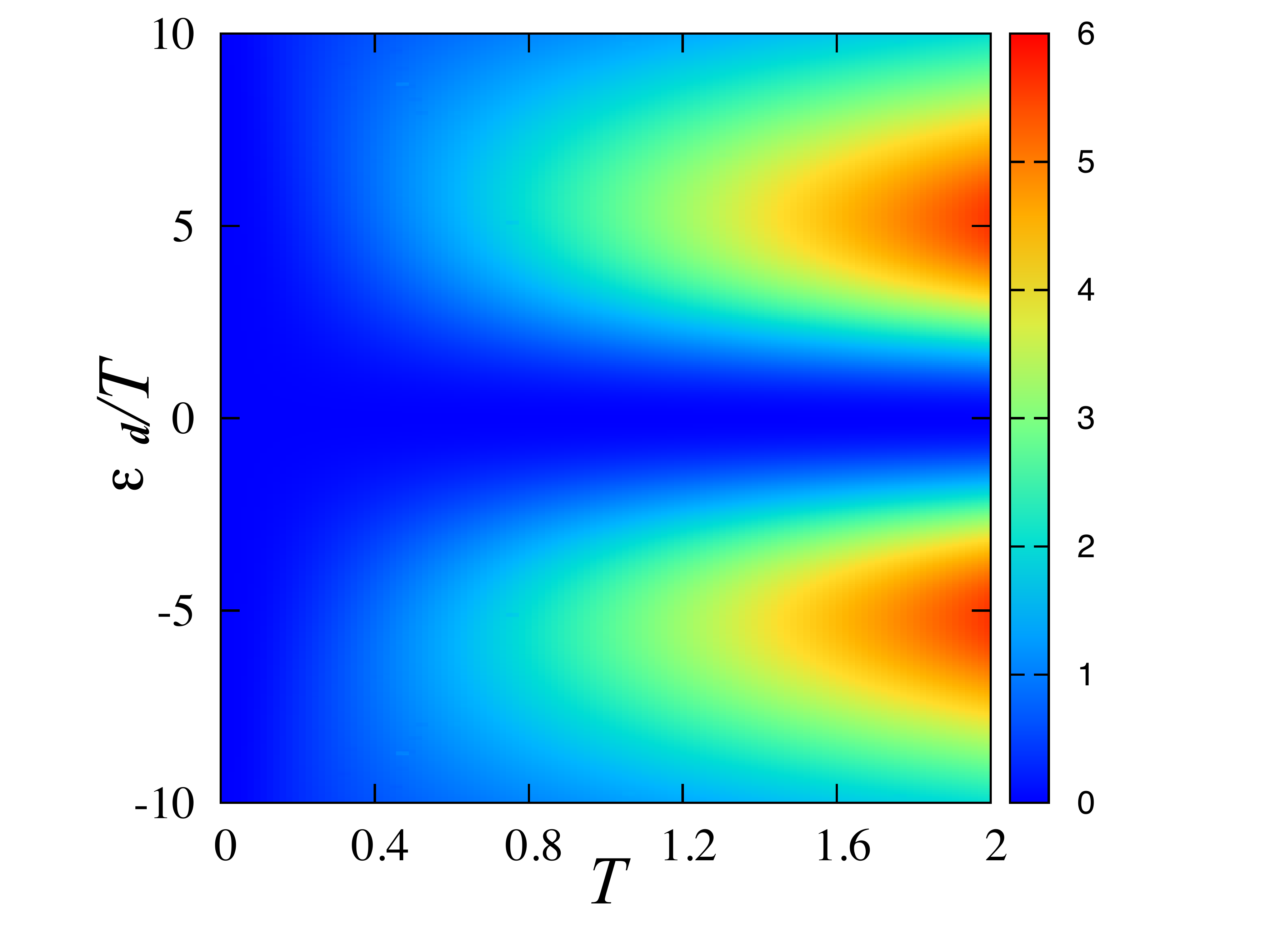}}
\caption{(Color online) Figure of merit $ZT$ as a function of the temperature $T$ and the energy of the QD level $\varepsilon_d$ (right panel). Plots at fixed temperatures are shown in the left panels.
Other details are the same as in Fig. \ref{fig2}.} 
\label{zt}
\end{figure}

{\em Figure of merit}. The response described by the Onsager coefficients is suppressed for fractional fillings. However, 
as the  off-diagonal elements are less suppressed than that of the diagonal ones, the thermoelectric response can be improved. An indication of the degree of such improvement is 
quantified by the figure of merit $ZT$. The corresponding behavior is shown in Fig. \ref{zt}. At all temperatures, there is an enhancement of the figure of merit in the fractional case relative to the integer one.
This effect is particularly crucial for low temperatures, where $ZT$ is very low in the integer QHE, but it can be improved up to an order of magnitude in the fractional one. 

The key for the understanding  of the behavior of the transport coefficients is the analysis of the transmission function given in Eq. (\ref{tau}). For simplicity, we focus on the case where the QD contains a single level, to study
the qualitative features of this function 
and the integrands of the Onsager coefficients.  Notice that in the limit 
$\Delta \gg  \gamma$  each of the levels of the QD contribute separately. For simplicity, we  focus on the case of a single level.  In the limit where $\gamma \rightarrow 0$, the density of states of the QD can be approximated as
$D_d(\varepsilon) \sim \delta \left(\varepsilon  - \varepsilon_d \right)$. Substituting in Eq. (\ref{l}) we have
\begin{equation} \label{l}
\hat{L}\sim -\frac{k_B T}{2 h}   \left(\begin{array}{cc} 
e & e \varepsilon_d \\
\varepsilon_d & \varepsilon_d^2 \end{array} \right) D_L(\varepsilon_d) D_R(\varepsilon_d) \;  f^{\prime}(\varepsilon_d), \;\;\;\;\;\;\; \gamma \rightarrow 0.
\end{equation}
Using these coefficients in the definition of the figure of merit, we find $ZT \rightarrow \infty$, which corresponds to Carnot efficiency. The fact that the limit of a vanishing width of the transmission function leads to optimal 
efficiency was originally pointed out in Ref. \onlinecite{mah-so} and constitutes an extreme case. In what follows we turn to analyze the case where $\gamma$ is a finite quantity, although it satisfies $\gamma \ll \Delta$.
For low energies, the density of states of the edge states behaves
as a power law, $D_{\alpha}(\varepsilon) \propto |\varepsilon|^{m_{\alpha}-1}$. On the other hand the difference of the Fermi functions sets the relevant integration window to 
$\sim [-k_B T,k_B T]$.  In the low-temperature regime $k_B T \leq \gamma$, we can approximate the density of states of the quantum dot by
$D_d(\varepsilon)=D_d(0)+D^{\prime}_d(0) \varepsilon$. 
The result of these approximations  leads to the following rough estimates of the Onsager matrix elements
\begin{eqnarray} \label{lowt}
L_{11} & \sim & c_{m_L,m_R}(T)  \frac{  D_d(0)  }{\left( \tilde{m}-1\right)} (k_B T)^{\tilde{m}-1}, \nonumber \\
L_{12}  &\sim & c_{m_L,m_R}(T)  \frac{D^{\prime}_d(0)  }{ \left( \tilde{m}+1 \right) } (k_B T)^{\tilde{m}+1}, \nonumber\\
L_{22} & \sim & c_{m_L,m_R}(T) \frac{ D_d(0)}{\left( \tilde{m}+1 \right)} (k_B T)^{\tilde{m}+1},
\end{eqnarray}
 where
the common prefactor $c_{m_L,m_R}(T)$ is a function of the temperature. As already  stressed before,  the low-temperature behavior of these coefficients is determined by $\tilde{m}=m_L+m_R$.

We can now see that the Seebeck coefficient is approximately given by
\begin{equation}\label{apro}
S \simeq -  2  \; \frac{\left( \tilde{m}-1 \right)}{\left( \tilde{m}+1\right)} \; \frac{\varepsilon_d \;k_B T }{\varepsilon_d ^2 + \gamma^2}, \;\;\;\;\;\;\;\; T <\gamma.
\end{equation}
This implies an enhancement of at least a factor $ 3  \left( \tilde{m}-1 \right)/\left( \tilde{m}+1\right)$, relative to the integer-filling case, which is equivalent to $6 /5$ for $\tilde{m}=4$, corresponding to 
filling factors $(\nu_L,\nu_R)=(1, 1/3)$ or $ (1/3, 1)$,  and $15 /7$ for  $\tilde{m}=6$, corresponding to $(\nu_L,\nu_R)=(1/3,1/3)$. Similarly, the figure of merit can be written as 
\begin{equation}\label{aprozt}
ZT= \frac{1}{\alpha -1}, \;\;\;\;\;\;\;\;\;\; \alpha= \frac{\left(\varepsilon_d^2 + \gamma^2 \right)^2}{\left( 2 k_B T \varepsilon_d  \right)^2} 
\frac{\left( \tilde{m}+1 \right)}{\left( \tilde{m}-1\right) },
\end{equation}
with $ \alpha= L_{11} L_{22}/ L_{12}^2$. Concomitant with the behavior of the Seebeck coefficient, we see that as $\tilde{m}$ increases $\alpha \rightarrow 1$, which implies an 
enhancement of the figure of merit, which is, of course, restricted to the  limited number of physically realized values of $\tilde{m}$. 
We have verified that these simple approximate expressions fit on top of the exact results for temperatures $0<k_B T<0.1 \gamma$. For higher temperatures these expressions
remain qualitatively correct while quantitatively lower than the exact result. Hence,  Eqs. (\ref{apro}) and (\ref{aprozt}) provide quite accurate lower bounds
 for the exact thermoelectric performance of the device of Fig. \ref{fig1}. It is important to stress that in the case of the figure of merit $ZT$ the enhancement in the fractional case
 increases significantly as the temperature grows.

In order to gather the relevance of these results in the context of concrete experimental situations, let us quote the typical energy scales for quantum dots embedded in quantum Hall bars. 
The value of the level spacing $\Delta$ varies  in different experimental setups. It can be of the order of $200 \mu eV$ \cite{feve}. In Ref. \cite{altimiras2} it is quoted $32 \mu eV$. Assuming that  typical values for the hybridization energy are $\gamma \sim 1 \mu eV$ and considering  the values of $\Delta$ mentioned above,
the behavior of the transport coefficients can be regarded  as a trivial 
superposition of the contributions of the single levels.  
For these estimates, the  temperature of the left middle panel of Fig. \ref{zt} corresponds to
 $T \sim 12 mK$.
  For these parameters, the maximum value of the figure of merit is  $ZT  \sim 1$ in the integer case, 
which implies $\eta < \eta_C/6$. Instead, for  filling factors $(\nu_L,\nu_R)=(1/3,1/3)$ the figure of merit may achieve $ZT \sim 3$, implying $\eta \sim \eta_C/3$ at the maxima. For higher temperatures, 
like the one
 shown in the upper left panel of the figure (corresponding to  $T \sim 60 mK$), $ZT \sim 15$, at the maxima in the case with $(\nu_L,\nu_R)=(1/3,1/3)$, 
implying $\eta \sim 3/5 \eta_C$. 

{\em Conclusions}. We have investigated the thermoelectric performance of a two-terminal quantum Hall effect device with an embedded quantum dot. We have shown that the thermoelectric response 
described by the Seebeck coefficient, as well as the figure of merit which parametrizes the maximum efficiency, increase for increasing values of the inverse of the filling factor. Estimates of the parameters involved
indicate that 
this effect should be relevant in typical operating conditions of the quantum Hall effect.

 LA thanks  the support
of the Alexander von Humboldt Foundation.  EF thanks the hospitality and support of the Physics Dept. University of Buenos Aires and 
partial support from the USA National Science Foundation grant DMR 1725401 at the University of Illinois. 
PR and LA
acknowledge support from CONICET, MINCyT and UBACyT,
Argentina.

\newpage

\begin{widetext}

\begin{center}

{\bf \large Supplemental Material: Enhanced thermoelectric response in the fractional quantum Hall effect}

\end{center}

\onecolumngrid

\appendix
\pagestyle{plain}

\section{Schwinger-Keldysh approach and perturbative evaluation of the currents}\label{pert}

\textbf{\textit{Charge current}}

Starting from the average of the charge current operator and the definitions of the fermionic Green functions
$G^{<}_{\alpha, d}( t,t^{\prime})=i\langle d^{\dagger}_{\sigma}(t^{\prime})\Psi_{\alpha}(xt) \rangle$
and
$G_{\alpha,\alpha}^{\eta\eta'}(x, x^{\prime};t,t')=-i\langle \hat{T} (\Psi_{\alpha}(xt^{\eta})\Psi^{\dagger}_{\alpha}(x'{t'}^{\eta'}))
\rangle$,  
and\\
$G_{d,d}^{\eta\eta'}(t,t')=-i\langle \hat{T} (d_{\sigma}(t^{\eta})d_{\sigma}^{\dagger}({t'}^{\eta'}))
\rangle$
with $\eta=+,-$ labeling the upper and lower branches of the Keldysh contour respectively. 
For simplicity, we will omit the spacial coordinates of the fields, which correspond to 
$x=x^{\prime}=x_0$.

The non diagonal Green function $G^{<}_{L d}(t-t')$ obeys the Dyson expansion
and considering the lowest order approximation in $V_{t}$, $G_{i,i} \approx g_i$, it reads as follows
\begin{eqnarray}\label{lesser-Ld-1}
G^{<}_{L d}(t-t')& =& V_{t}\int_{-\infty}^{+\infty} dt_{1}\Bigr( g_{L}^{++}(t,t_{1}) g_{d}^{+-}(t_{1},t') \nonumber \\
                                                   & &        - g_{L}^{+-}(t,t_{1}) g_{d}^{--}(t_{1},t')\Bigl).
\end{eqnarray}

In order to include the chemical potential, it is convenient to introduce the following gauge transformation 
in the fermionic fields describing the lead $\alpha$

\begin{eqnarray}\label{gaige}
\Psi_{\alpha}(t)&&\rightarrow ~e^{it\mu_{\alpha}}\Psi_{\alpha}(t),
\end{eqnarray}

which yields the following transformation of the Green function
$g_{\alpha}^{\eta\eta'}(t)\rightarrow e^{it\mu_{\alpha}}g_{\alpha}^{\eta\eta'}(t)$. 
Therefore Eq. (\ref{lesser-Ld-1}) at equal times reads

\begin{eqnarray}\label{lesser-RL-3}
G^{<}_{L d}(t-t')\vert_{t'=t}&=&V_{t}\int_{-\infty}^{+\infty} dt_{1}e^{i(t-t_{1})\mu_{L}}
\Bigr( g_{L}^{++}(t-t_{1}) g_{d}^{+-}(t_{1}-t) \nonumber\\
 &&- g_{L}^{+-}(t-t_{1}) g_{d}^{--}(t_{1}-t)\Bigl).
\end{eqnarray}

By using the following identity for the different components of the Green function along the Keldysh contour,
\begin{equation}
 g_{i}^{--}=g_{i}^{+-}+g_{i}^{-+}-g_{i}^{++},
 \end{equation}
and the notation $g_{i}^{+-}=g_{i}^{<}$, $g_{i}^{-+}=g_{i}^{>}$,  the charge current becomes,
\begin{eqnarray}\label{J-current-6}
& & J_{C} = eV_{t} \Bigl( G^{<}_{d,L}(t-t')\vert_{t'=t}-G^{<}_{L,d}(t-t')\vert_{t'=t}\Bigl)\\ 
     &=& eV^{2}_{t}\int_{-\infty}^{+\infty} dte^{it \mu_L}
\Bigr( g_{L}^{<}(t) g_{d}^{>}(-t) - g_{L}^{>}(t) g_{d}^{<}(-t)\Bigl),\nonumber
\end{eqnarray}

and introducing the bosonic representation to express the fermionic Green function in terms of the bosonic one
\begin{equation} \label{boson}
g_{\alpha}^{\lessgtr}(t)=\pm\frac{i}{2\pi a}e^{\nu^{-2} D_{\alpha}^{\lessgtr}(t)},
\end{equation}
where $D_{\alpha}^{\eta\eta'}(t)$ are the components along the Keldysh contour
of the bosonic Green function \cite{martin},

\begin{eqnarray} \label{bosgreen}
D_{\alpha}^{\eta\eta'}(t,t')&=&
              \langle \hat{T} \left( \phi_{\alpha}(t^{\eta})\phi_{\alpha}(t'^{\eta'}) \right) \rangle 
              -\frac{\langle\phi_{\alpha}(t^{\eta})\rangle^{2}}{2}
              -\frac{\langle\phi_{\alpha}(t'^{\eta'})\rangle^{2}}{2}\\\nonumber
D_{\alpha}^{\eta,-\eta}(t)&=&- \nu \ln \Big( \sinh \big( \pi T_{\alpha} (\eta t + ia) \big) / 
                                        \sinh \big( i\pi T_{\alpha} a \big) \Big)           
\end{eqnarray}

Notice that the relation $D^{+-}(t)=D^{-+}(-t)$ or $D^{<}(t)=D^{>}(-t)$, explicitly takes into account
the particle-hole symmetry of each lead.

It is convenient to introduce the Fourier transform of the lesser Green functions describing the leads, 
\begin{eqnarray}\label{g-lesser}
g_{\alpha}^{<}(t) &=& \frac{i}{2\pi a} ~\frac{\sinh^{m_{\alpha}}(ia\pi T_{\alpha})}
                                             {\sinh^{m_{\alpha}}[\pi T_{\alpha}(t+ia)]}\\
g_{\alpha}^{<}(\varepsilon) &=& \frac{i}{2\pi a} ~ a^{m_{\alpha}}
                       \frac{(2\pi T_{\alpha})^{{m_{\alpha}}-1}}{\Gamma({m_{\alpha}})}
                       e^{-\varepsilon/2T_{\alpha}}\Big\vert \Gamma[{m_{\alpha}}/2 + i\varepsilon/(2\pi T_{\alpha})]\Big\vert ^2.
\end{eqnarray}

Furthermore, with the help of the following identity for the Fermi function,  
$f(\varepsilon)= \big(  e^{~\varepsilon/T} + 1 \big)^{-1} = 
\frac{1}{2\pi}e^{-\varepsilon/2T}\vert \Gamma[1/2 + i\varepsilon/(2\pi T)] \vert ^2$, 
the lesser Green function $g_{\alpha}^{<}(\varepsilon)$ can be written in a more familiar form, 
$g_{\alpha}^{<}(\varepsilon)=2\pi i D_{\alpha} (\varepsilon)   f_{\alpha}(\varepsilon)$,\cite{bra} where we have introduced the 
spectral function of the lead $\alpha$

\begin{eqnarray}\label{rho_alpha}
 D_{\alpha} (\varepsilon)=a^{{m_{\alpha}}-1} \frac{(2\pi T_{\alpha})^{{m_{\alpha}}-1}}{2\pi\Gamma({m_{\alpha}})}
                       \frac{\vert \Gamma[{m_{\alpha}}/2 + i\varepsilon/(2\pi T_{\alpha})] \vert ^2}
                            {\vert \Gamma[1/2 + i\varepsilon/(2\pi T)] \vert ^2}.
\end{eqnarray}

On the other hand, the Green function for  the dot, evaluated at ${\cal O}(V_t^2)$ reads
\begin{equation}
g_d^{<,>}(\varepsilon)=  
V_t^2 |g_d^{r}(\varepsilon)|^2 \left[\sum_{\alpha=L,R} g_{\alpha}^{<,>}(\varepsilon) \right] = \frac{V_t^2}{\gamma} D_d(\varepsilon) \left[\sum_{\alpha=L,R} g_{\alpha}^{<,>}(\varepsilon) \right],
\end{equation}
where 
\begin{equation}
D_d(\varepsilon) = |g_d^{r}(\varepsilon)|^2 \gamma= \sum_j \frac{\gamma/N\pi}
{\left(\varepsilon- \varepsilon_{d,j} \right)^2 + \gamma^2}
\end{equation}
 is the spectral density of the quantum dot, which we model as   $N$ resonant levels with energies $\varepsilon_{d,j}, \; j= 1, \ldots N$ and widths $\gamma$.

Substituting these expressions in Eq. (\ref{J-current-6}) we get
\begin{eqnarray}\label{j-final}
J_C = \frac{e}{h} \int d \varepsilon \; \tau(\varepsilon)\;  \left[ f_L(\varepsilon+\mu_L)- f_R(\varepsilon+\mu_R) \right],
\end{eqnarray}

where we have defined the transmission function 
\begin{equation} \label{tau}
\tau(\varepsilon)= \frac{V_t^4}{\gamma}\; D_R(\varepsilon + \mu_R) \; D_d(\varepsilon)\;  D_L(\varepsilon + \mu_L).
\end{equation}

\vspace{1cm}
\textbf{\textit{Energy current}}

Regarding the energy current, $J_E$, some care should be taken when introducing the chemical potentials through 
the gauge transformation in Eq.(\ref{gaige}). Starting from Eq.(6) of the main text and taking in mind that
$\dot{g}_{\alpha}(t)\rightarrow i\mu_{\alpha}e^{i\mu_{\alpha}t}g_{\alpha}(t)+e^{i\mu_{\alpha}t}\dot{g}_{\alpha}(t)$, 
after applying  Eq.\ref{gaige}, with $\dot{a}(t)=\partial_t a(t)$, and performing the same perturbative approximation in 
Eq. (\ref{lesser-RL-3}) 
we have
\begin{eqnarray}\label{j-energy-1}
J_E&=& -\frac{\mu_L}{e} J_{C_L} +  eV^{2}_{t}\int_{-\infty}^{+\infty} dt~e^{it \mu_L}
\Bigr( \dot{g}_{L}^{<}(t) g_{d}^{>}(-t) - \dot{g}_{L}^{>}(t) g_{d}^{<}(-t)\Bigl)                      
\end{eqnarray}

Note that by using the energy current conservation, $J_E=J_{E_L}=-J_{E_R}$, the first term in Eq.(\ref{j-energy-1})
vanishes due to the fact that $-\frac{\mu_L}{e} J_{C_L}+\frac{\mu_R}{e} J_{C_R}=
\frac{\mu_L+\mu_R}{e} J_{C_R}=0$ and $\mu_L+\mu_R=0$.
After that and following the same steps sketch in the charge current, the final expression of the energy one can 
by written in the form

\begin{eqnarray}\label{je-final}
J_E = \frac{e}{h} \int d \varepsilon \; \varepsilon\; \tau(\varepsilon)\;  \left[ f_L(\varepsilon+\mu_L)- f_R(\varepsilon+\mu_R) \right],
\end{eqnarray}

\end{widetext}

\end{document}